\documentclass[journal,10pt,twoside,twocolumn]{IEEEtran}
\usepackage{cite}
\usepackage[T1]{fontenc}
\usepackage{graphicx}
\usepackage{amssymb}
\usepackage{amsmath}
\usepackage{paralist}
\usepackage{subfigure}
\usepackage{booktabs} %\toprule \midrule \bottomrule
\usepackage{algorithm}
\usepackage{algorithmic}
\usepackage{amsthm}
\usepackage{multirow}
\usepackage{microtype}
\usepackage{tablefootnote}

\usepackage{amssymb}
\usepackage{amsfonts}
\usepackage{mathrsfs}
\usepackage{xspace}
\usepackage{bm}
\usepackage{upgreek}

\newcommand{\safemath}[2]{\newcommand{#1}{\ensuremath{#2}\xspace}}

\safemath{\bma}{\mathbf{a}}
\safemath{\bmb}{\mathbf{b}}
\safemath{\bmc}{\mathbf{c}}
\safemath{\bmd}{\mathbf{d}}
\safemath{\bme}{\mathbf{e}}
\safemath{\bmf}{\mathbf{f}}
\safemath{\bmg}{\mathbf{g}}
\safemath{\bmh}{\mathbf{h}}
\safemath{\bmi}{\mathbf{i}}
\safemath{\bmj}{\mathbf{j}}
\safemath{\bmk}{\mathbf{k}}
\safemath{\bml}{\mathbf{l}}
\safemath{\bmm}{\mathbf{m}}
\safemath{\bmn}{\mathbf{n}}
\safemath{\bmo}{\mathbf{o}}
\safemath{\bmp}{\mathbf{p}}
\safemath{\bmq}{\mathbf{q}}
\safemath{\bmr}{\mathbf{r}}
\safemath{\bms}{\mathbf{s}}
\safemath{\bmt}{\mathbf{t}}
\safemath{\bmu}{\mathbf{u}}
\safemath{\bmv}{\mathbf{v}}
\safemath{\bmw}{\mathbf{w}}
\safemath{\bmx}{\mathbf{x}}
\safemath{\bmy}{\mathbf{y}}
\safemath{\bmz}{\mathbf{z}}
\safemath{\bmzero}{\mathbf{0}}
\safemath{\bmone}{\mathbf{1}}

\bmdefine{\biad}{a}
\bmdefine{\bibd}{b}
\bmdefine{\bicd}{c}
\bmdefine{\bidd}{d}
\bmdefine{\bied}{e}
\bmdefine{\bifd}{f}
\bmdefine{\bigd}{g}
\bmdefine{\bihd}{h}
\bmdefine{\biid}{i}
\bmdefine{\bijd}{j}
\bmdefine{\bikd}{k}
\bmdefine{\bild}{l}
\bmdefine{\bimd}{m}
\bmdefine{\bind}{n}
\bmdefine{\biod}{o}
\bmdefine{\bipd}{p}
\bmdefine{\biqd}{q}
\bmdefine{\bird}{r}
\bmdefine{\bisd}{s}
\bmdefine{\bitd}{t}
\bmdefine{\biud}{u}
\bmdefine{\bivd}{v}
\bmdefine{\biwd}{w}
\bmdefine{\bixd}{x}
\bmdefine{\biyd}{y}
\bmdefine{\bizd}{z}

\bmdefine{\bixid}{\xi}
\bmdefine{\bilambdad}{\lambda}
\bmdefine{\bimud}{\mu}
\bmdefine{\bithetad}{\theta}
\bmdefine{\biphid}{\phi}
\bmdefine{\bideltad}{\delta}

\safemath{\bmia}{\biad}
\safemath{\bmib}{\bibd}
\safemath{\bmic}{\bicd}
\safemath{\bmid}{\bidd}
\safemath{\bmie}{\bied}
\safemath{\bmif}{\bifd}
\safemath{\bmig}{\bigd}
\safemath{\bmih}{\bihd}
\safemath{\bmii}{\biid}
\safemath{\bmij}{\bijd}
\safemath{\bmik}{\bikd}
\safemath{\bmil}{\bild}
\safemath{\bmim}{\bimd}
\safemath{\bmin}{\bind}
\safemath{\bmio}{\biod}
\safemath{\bmip}{\bipd}
\safemath{\bmiq}{\biqd}
\safemath{\bmir}{\bird}
\safemath{\bmis}{\bisd}
\safemath{\bmit}{\bitd}
\safemath{\bmiu}{\biud}
\safemath{\bmiv}{\bivd}
\safemath{\bmiw}{\biwd}
\safemath{\bmix}{\bixd}
\safemath{\bmiy}{\biyd}
\safemath{\bmiz}{\bizd}

\safemath{\bmxi}{\bixid}
\safemath{\bmlambda}{\bilambdad}
\safemath{\bmmu}{\bimud}
\safemath{\bmtheta}{\bithetad}
\safemath{\bmphi}{\biphid}
\safemath{\bmdelta}{\bideltad}

\safemath{\bA}{\mathbf{A}}
\safemath{\bB}{\mathbf{B}}
\safemath{\bC}{\mathbf{C}}
\safemath{\bD}{\mathbf{D}}
\safemath{\bE}{\mathbf{E}}
\safemath{\bF}{\mathbf{F}}
\safemath{\bG}{\mathbf{G}}
\safemath{\bH}{\mathbf{H}}
\safemath{\bI}{\mathbf{I}}
\safemath{\bJ}{\mathbf{J}}
\safemath{\bK}{\mathbf{K}}
\safemath{\bL}{\mathbf{L}}
\safemath{\bM}{\mathbf{M}}
\safemath{\bN}{\mathbf{N}}
\safemath{\bO}{\mathbf{O}}
\safemath{\bP}{\mathbf{P}}
\safemath{\bQ}{\mathbf{Q}}
\safemath{\bR}{\mathbf{R}}
\safemath{\bS}{\mathbf{S}}
\safemath{\bT}{\mathbf{T}}
\safemath{\bU}{\mathbf{U}}
\safemath{\bV}{\mathbf{V}}
\safemath{\bW}{\mathbf{W}}
\safemath{\bX}{\mathbf{X}}
\safemath{\bY}{\mathbf{Y}}
\safemath{\bZ}{\mathbf{Z}}

\safemath{\bZero}{\mathbf{0}}
\safemath{\bOne}{\mathbf{1}}
\safemath{\bDelta}{\mathbf{\Delta}}
\safemath{\bLambda}{\mathbf{\UpLambda}}
\safemath{\bPhi}{\mathbf{\Upphi}}
\safemath{\bSigma}{\mathbf{\Upsigma}}
\safemath{\bOmega}{\mathbf{\Upomega}}
\safemath{\bTheta}{\mathbf{\Uptheta}}

\bmdefine{\biAd}{A}
\bmdefine{\biBd}{B}
\bmdefine{\biCd}{C}
\bmdefine{\biDd}{D}
\bmdefine{\biEd}{E}
\bmdefine{\biFd}{F}
\bmdefine{\biGd}{G}
\bmdefine{\biHd}{H}
\bmdefine{\biId}{I}
\bmdefine{\biJd}{J}
\bmdefine{\biKd}{K}
\bmdefine{\biLd}{L}
\bmdefine{\biMd}{M}
\bmdefine{\biOd}{N}
\bmdefine{\biPd}{O}
\bmdefine{\biQd}{P}
\bmdefine{\biRd}{R}
\bmdefine{\biSd}{S}
\bmdefine{\biTd}{T}
\bmdefine{\biUd}{U}
\bmdefine{\biVd}{V}
\bmdefine{\biWd}{W}
\bmdefine{\biXd}{X}
\bmdefine{\biYd}{Y}
\bmdefine{\biZd}{Z}

\bmdefine{\biDelta}{\Delta}
\bmdefine{\biLambda}{\Lambda}
\bmdefine{\biPhi}{\Phi}
\bmdefine{\biSigma}{\Sigma}
\bmdefine{\biOmega}{\Omega}
\bmdefine{\biTheta}{\Theta}

\safemath{\bimA}{\biAd}
\safemath{\bimB}{\biBd}
\safemath{\bimC}{\biCd}
\safemath{\bimD}{\biDd}
\safemath{\bimE}{\biEd}
\safemath{\bimF}{\biFd}
\safemath{\bimG}{\biGd}
\safemath{\bimH}{\biHd}
\safemath{\bimI}{\biId}
\safemath{\bimJ}{\biJd}
\safemath{\bimK}{\biKd}
\safemath{\bimL}{\biLd}
\safemath{\bimM}{\biMd}
\safemath{\bimN}{\biNd}
\safemath{\bimO}{\biOd}
\safemath{\bimP}{\biPd}
\safemath{\bimQ}{\biQd}
\safemath{\bimR}{\biRd}
\safemath{\bimS}{\biSd}
\safemath{\bimT}{\biTd}
\safemath{\bimU}{\biUd}
\safemath{\bimV}{\biVd}
\safemath{\bimW}{\biWd}
\safemath{\bimX}{\biXd}
\safemath{\bimY}{\biYd}
\safemath{\bimZ}{\biZd}

\safemath{\bimDelta}{\biDelta}
\safemath{\bimLambda}{\biLambda}
\safemath{\bimPhi}{\biPhi}
\safemath{\bimSigma}{\biSigma}
\safemath{\bimOmega}{\biOmega}
\safemath{\bimTheta}{\biTheta}

\safemath{\setA}{\mathcal{A}}
\safemath{\setB}{\mathcal{B}}
\safemath{\setC}{\mathcal{C}}
\safemath{\setD}{\mathcal{D}}
\safemath{\setE}{\mathcal{E}}
\safemath{\setF}{\mathcal{F}}
\safemath{\setG}{\mathcal{G}}
\safemath{\setH}{\mathcal{H}}
\safemath{\setI}{\mathcal{I}}
\safemath{\setJ}{\mathcal{J}}
\safemath{\setK}{\mathcal{K}}
\safemath{\setL}{\mathcal{L}}
\safemath{\setM}{\mathcal{M}}
\safemath{\setN}{\mathcal{N}}
\safemath{\setO}{\mathcal{O}}
\safemath{\setP}{\mathcal{P}}
\safemath{\setQ}{\mathcal{Q}}
\safemath{\setR}{\mathcal{R}}
\safemath{\setS}{\mathcal{S}}
\safemath{\setT}{\mathcal{T}}
\safemath{\setU}{\mathcal{U}}
\safemath{\setV}{\mathcal{V}}
\safemath{\setW}{\mathcal{W}}
\safemath{\setX}{\mathcal{X}}
\safemath{\setY}{\mathcal{Y}}
\safemath{\setZ}{\mathcal{Z}}
\safemath{\emptySet}{\varnothing}

\safemath{\colA}{\mathscr{A}}
\safemath{\colB}{\mathscr{B}}
\safemath{\colC}{\mathscr{C}}
\safemath{\colD}{\mathscr{D}}
\safemath{\colE}{\mathscr{E}}
\safemath{\colF}{\mathscr{F}}
\safemath{\colG}{\mathscr{G}}
\safemath{\colH}{\mathscr{H}}
\safemath{\colI}{\mathscr{I}}
\safemath{\colJ}{\mathscr{J}}
\safemath{\colK}{\mathscr{K}}
\safemath{\colL}{\mathscr{L}}
\safemath{\colM}{\mathscr{M}}
\safemath{\colN}{\mathscr{N}}
\safemath{\colO}{\mathscr{O}}
\safemath{\colP}{\mathscr{P}}
\safemath{\colQ}{\mathscr{Q}}
\safemath{\colR}{\mathscr{R}}
\safemath{\colS}{\mathscr{S}}
\safemath{\colT}{\mathscr{T}}
\safemath{\colU}{\mathscr{U}}
\safemath{\colV}{\mathscr{V}}
\safemath{\colW}{\mathscr{W}}
\safemath{\colX}{\mathscr{X}}
\safemath{\colY}{\mathscr{Y}}
\safemath{\colZ}{\mathscr{Z}}

\safemath{\opA}{\mathbb{A}}
\safemath{\opB}{\mathbb{B}}
\safemath{\opC}{\mathbb{C}}
\safemath{\opD}{\mathbb{D}}
\safemath{\opE}{\mathbb{E}}
\safemath{\opF}{\mathbb{F}}
\safemath{\opG}{\mathbb{G}}
\safemath{\opH}{\mathbb{H}}
\safemath{\opI}{\mathbb{I}}
\safemath{\opJ}{\mathbb{J}}
\safemath{\opK}{\mathbb{K}}
\safemath{\opL}{\mathbb{L}}
\safemath{\opM}{\mathbb{M}}
\safemath{\opN}{\mathbb{N}}
\safemath{\opO}{\mathbb{O}}
\safemath{\opP}{\mathbb{P}}
\safemath{\opQ}{\mathbb{Q}}
\safemath{\opR}{\mathbb{R}}
\safemath{\opS}{\mathbb{S}}
\safemath{\opT}{\mathbb{T}}
\safemath{\opU}{\mathbb{U}}
\safemath{\opV}{\mathbb{V}}
\safemath{\opW}{\mathbb{W}}
\safemath{\opX}{\mathbb{X}}
\safemath{\opY}{\mathbb{Y}}
\safemath{\opZ}{\mathbb{Z}}
\safemath{\opZero}{\mathbb{O}}
\safemath{\identityop}{\opI}

\safemath{\veca}{\bma}
\safemath{\vecb}{\bmb}
\safemath{\vecc}{\bmc}
\safemath{\vecd}{\bmd}
\safemath{\vece}{\bme}
\safemath{\vecf}{\bmf}
\safemath{\vecg}{\bmg}
\safemath{\vech}{\bmh}
\safemath{\veci}{\bmi}
\safemath{\vecj}{\bmj}
\safemath{\veck}{\bmk}
\safemath{\vecl}{\bml}
\safemath{\vecm}{\bmm}
\safemath{\vecn}{\bmn}
\safemath{\veco}{\bmo}
\safemath{\vecp}{\bmp}
\safemath{\vecq}{\bmq}
\safemath{\vecr}{\bmr}
\safemath{\vecs}{\bms}
\safemath{\vect}{\bmt}
\safemath{\vecu}{\bmu}
\safemath{\vecv}{\bmv}
\safemath{\vecw}{\bmw}
\safemath{\vecx}{\bmx}
\safemath{\vecy}{\bmy}
\safemath{\vecz}{\bmz}

\safemath{\veczero}{\bmzero}
\safemath{\vecone}{\bmone}
\safemath{\vecxi}{\bmxi}
\safemath{\veclambda}{\bmlambda}
\safemath{\vecmu}{\bmmu}
\safemath{\vectheta}{\bmtheta}
\safemath{\vecphi}{\bmphi}
\safemath{\vecdelta}{\bmdelta}

\safemath{\matA}{\bA}
\safemath{\matB}{\bB}
\safemath{\matC}{\bC}
\safemath{\matD}{\bD}
\safemath{\matE}{\bE}
\safemath{\matF}{\bF}
\safemath{\matG}{\bG}
\safemath{\matH}{\bH}
\safemath{\matI}{\bI}
\safemath{\matJ}{\bJ}
\safemath{\matK}{\bK}
\safemath{\matL}{\bL}
\safemath{\matM}{\bM}
\safemath{\matN}{\bN}
\safemath{\matO}{\bO}
\safemath{\matP}{\bP}
\safemath{\matQ}{\bQ}
\safemath{\matR}{\bR}
\safemath{\matS}{\bS}
\safemath{\matT}{\bT}
\safemath{\matU}{\bU}
\safemath{\matV}{\bV}
\safemath{\matW}{\bW}
\safemath{\matX}{\bX}
\safemath{\matY}{\bY}
\safemath{\matZ}{\bZ}
\safemath{\matzero}{\bmzero}

\safemath{\matDelta}{\bDelta}
\safemath{\matLambda}{\bLambda}
\safemath{\matPhi}{\bPhi}
\safemath{\matSigma}{\bSigma}
\safemath{\matOmega}{\bOmega}
\safemath{\matTheta}{\bTheta}

\safemath{\matidentity}{\matI}
\safemath{\matone}{\matO}

\safemath{\rnda}{A}
\safemath{\rndb}{B}
\safemath{\rndc}{C}
\safemath{\rndd}{D}
\safemath{\rnde}{E}
\safemath{\rndf}{F}
\safemath{\rndg}{G}
\safemath{\rndh}{H}
\safemath{\rndi}{I}
\safemath{\rndj}{J}
\safemath{\rndk}{K}
\safemath{\rndl}{L}
\safemath{\rndm}{M}
\safemath{\rndn}{N}
\safemath{\rndo}{O}
\safemath{\rndp}{P}
\safemath{\rndq}{Q}
\safemath{\rndr}{R}
\safemath{\rnds}{S}
\safemath{\rndt}{T}
\safemath{\rndu}{U}
\safemath{\rndv}{V}
\safemath{\rndw}{W}
\safemath{\rndx}{X}
\safemath{\rndy}{Y}
\safemath{\rndz}{Z}

\safemath{\rveca}{\bimA}
\safemath{\rvecb}{\bimB}
\safemath{\rvecc}{\bimC}
\safemath{\rvecd}{\bimD}
\safemath{\rvece}{\bimE}
\safemath{\rvecf}{\bimF}
\safemath{\rvecg}{\bimG}
\safemath{\rvech}{\bimH}
\safemath{\rveci}{\bimI}
\safemath{\rvecj}{\bimJ}
\safemath{\rveck}{\bimK}
\safemath{\rvecl}{\bimL}
\safemath{\rvecm}{\bimM}
\safemath{\rvecn}{\bimN}
\safemath{\rveco}{\bomO}
\safemath{\rvecp}{\bimP}
\safemath{\rvecq}{\bimQ}
\safemath{\rvecr}{\bimR}
\safemath{\rvecs}{\bimS}
\safemath{\rvect}{\bimT}
\safemath{\rvecu}{\bimU}
\safemath{\rvecv}{\bimV}
\safemath{\rvecw}{\bimW}
\safemath{\rvecx}{\bimX}
\safemath{\rvecy}{\bimY}
\safemath{\rvecz}{\bimZ}

\safemath{\rvecxi}{\bmxi}
\safemath{\rveclambda}{\bmlambda}
\safemath{\rvecmu}{\bmmu}
\safemath{\rvectheta}{\bmtheta}
\safemath{\rvecphi}{\bmphi}

\safemath{\rmatA}{\bimA}
\safemath{\rmatB}{\bimB}
\safemath{\rmatC}{\bimC}
\safemath{\rmatD}{\bimD}
\safemath{\rmatE}{\bimE}
\safemath{\rmatF}{\bimF}
\safemath{\rmatG}{\bimG}
\safemath{\rmatH}{\bimH}
\safemath{\rmatI}{\bimI}
\safemath{\rmatJ}{\bimJ}
\safemath{\rmatK}{\bimK}
\safemath{\rmatL}{\bimL}
\safemath{\rmatM}{\bimM}
\safemath{\rmatN}{\bimN}
\safemath{\rmatO}{\bimO}
\safemath{\rmatP}{\bimP}
\safemath{\rmatQ}{\bimQ}
\safemath{\rmatR}{\bimR}
\safemath{\rmatS}{\bimS}
\safemath{\rmatT}{\bimT}
\safemath{\rmatU}{\bimU}
\safemath{\rmatV}{\bimV}
\safemath{\rmatW}{\bimW}
\safemath{\rmatX}{\bimX}
\safemath{\rmatY}{\bimY}
\safemath{\rmatZ}{\bimZ}

\safemath{\rmatDelta}{\bimDelta}
\safemath{\rmatLambda}{\bimLambda}
\safemath{\rmatPhi}{\bimPhi}
\safemath{\rmatSigma}{\bimSigma}
\safemath{\rmatOmega}{\bimOmega}
\safemath{\rmatTheta}{\bimTheta}

\usepackage{amssymb}
\usepackage{amsfonts}
\usepackage{mathrsfs}
\usepackage{xspace}
\usepackage{bm}
\usepackage{fancyref}
\usepackage{textcomp}

\usepackage{multirow}
\usepackage{stmaryrd}

\newenvironment{textbmatrix}{	\setlength{\arraycolsep}{2.5pt}%
								\big[\begin{matrix}}{\end{matrix}\big]%
								\raisebox{0.08ex}{\vphantom{M}}}

\def\be{\begin{equation}}
\def\ee{\end{equation}}
\def\een{\nonumber \end{equation}}
\def\mat{\begin{bmatrix}}
\def\emat{\end{bmatrix}}
\def\btm{\begin{textbmatrix}}
\def\etm{\end{textbmatrix}}

\def\ba#1\ea{\begin{align}#1\end{align}}
\def\bas#1\eas{\begin{align*}#1\end{align*}}
\def\bs#1\es{\begin{split}#1\end{split}}
\def\bg#1\eg{\begin{gather}#1\end{gather}}
\def\bml#1\eml{\begin{multline}#1\end{multline}}
\def\bi#1\ei{\begin{itemize}#1\end{itemize}}

\newcommand{\lefto}{\mathopen{}\left}

\DeclareMathOperator*{\argmin}{arg\;min}		% arg min
\DeclareMathOperator{\Exop}{\opE}			% expectation operator
\DeclareMathOperator{\Varop}{\opV\!\mathrm{ar}} % variance operator
\newcommand{\Ex}[2]{\ensuremath{\Exop_{#1}\lefto[#2\right]}} 	% expectation
\newcommand{\abs}[1]{\lefto\lvert#1\right\rvert}		% absolute value

\newcommand{\vecnorm}[1]{\lefto\lVert#1\right\rVert}		% vector norm
\safemath{\dirac}{\delta}					% Dirac delta
\safemath{\krond}{\dirac}					% Kronecker delta
\safemath{\upto}{\uparrow}
\safemath{\downto}{\downarrow}
\safemath{\iu}{j}							% imaginary unit
\safemath{\ev}{\lambda}						% eigenvalue
\safemath{\hilseqspace}{l^{2}}				% Hilbert sequence space
\newcommand{\banachfunspace}[1]{\setL^{#1}}	% Banach function space
\safemath{\hilfunspace}{\banachfunspace{2}}	% Hilbert function space
\safemath{\SNR}{\textit{SNR}} 				% signal to noise ratio
\safemath{\PAR}{\textit{PAR}} 				% signal to noise ratio
\safemath{\No}{N_0}							% noise spectral density
\safemath{\Es}{E_s}							% energy per symbol
\safemath{\Eb}{E_b}							% energy per bit
\safemath{\EbNo}{\frac{\Eb}{\No}}
\safemath{\EsNo}{\frac{\Es}{\No}}

\DeclareMathOperator{\CHop}{\ensuremath{\opH}} % channel operator
\safemath{\tvir}{\rndh_{\CHop}}				% time-varying impulse response
\safemath{\tvtf}{\rndl_{\CHop}}				% 	-''- transfer function
\safemath{\spf}{\rnds_{\CHop}}				% spreading function
\safemath{\bff}{H_{\CHop}}					% bi-freuqency function

\safemath{\ircf}{r_{h}}						% impulse response correlation fn.
\safemath{\tftvcf}{r_{s}}					% scattering function
\safemath{\tfcf}{r_{l}}						% time-frequency correlation fn.
\safemath{\bfcf}{r_{H}}						% bi-frequency correlation fn.

\safemath{\tcorr}{c_h}						% time-correlation function
\safemath{\scf}{c_{s}}						% spreading function
\safemath{\tfcorr}{c_{l}}					% transfer-function correlation
\safemath{\fcorr}{c_{H}}						% frequency-correlation function

\safemath{\mi}{I}							% mutual information
\safemath{\capacity}{C}						% capacity

\safemath{\normal}{\mathcal{N}}			% normal distribution
\safemath{\jpg}{\mathcal{CN}}			% jointly proper Gaussian
\safemath{\mchain}{\leftrightarrow}		% Markov chain
\safemath{\dB}{\,\mathrm{dB}}
\safemath{\dBm}{\,\mathrm{dBm}}
\safemath{\Hz}{\,\mathrm{Hz}}
\safemath{\kHz}{\,\mathrm{kHz}}
\safemath{\MHz}{\,\mathrm{MHz}}
\safemath{\GHz}{\,\mathrm{GHz}}
\safemath{\s}{\,\mathrm{s}}
\safemath{\ms}{\,\mathrm{ms}}
\safemath{\mus}{\,\mathrm{\text{\textmu}s}}
\safemath{\ns}{\,\mathrm{ns}}
\safemath{\ps}{\,\mathrm{ps}}
\safemath{\meter}{\,\mathrm{m}}
\safemath{\mm}{\,\mathrm{mm}}
\safemath{\cm}{\,\mathrm{cm}}
\safemath{\m}{\,\mathrm{m}}
\safemath{\W}{\,\mathrm{W}}
\safemath{\mW}{\, \mathrm{mW}}
\safemath{\J}{\,\mathrm{J}}
\safemath{\K}{\,\mathrm{K}}
\safemath{\bit}{\,\mathrm{bit}}
\safemath{\nat}{\,\mathrm{nat}}

\safemath{\define}{\triangleq}			% definition

\safemath{\equivalent}{\sim}
\safemath{\distas}{\sim}					% distributed according to
\safemath{\sdiff}{\Delta}				% symmetric set difference

\safemath{\reals}{\mathbb{R}}
\safemath{\positivereals}{\reals_{+}}
\safemath{\integers}{\mathbb{Z}}
\safemath{\posint}{\integers_{+}}
\safemath{\naturals}{\mathbb{N}}
\safemath{\posnaturals}{\naturals_{+}}
\safemath{\complexset}{\mathbb{C}}
\safemath{\rationals}{\mathbb{Q}}

\newcommand*{\fancyrefapplabelprefix}{app}		% Appendix
\newcommand*{\fancyrefthmlabelprefix}{thm}		% Theorem
\newcommand*{\fancyreflemlabelprefix}{lem}		% Lemma
\newcommand*{\fancyrefcorlabelprefix}{cor}		% Corollary
\newcommand*{\fancyrefdeflabelprefix}{def}		% Definition
\newcommand*{\fancyrefproplabelprefix}{prop}		% Proposition
\newcommand*{\fancyrefexmpllabelprefix}{exmpl}
\newcommand*{\fancyrefalglabelprefix}{alg}		% Algorithm
\newcommand*{\fancyreftbllabelprefix}{tbl}		% Algorithm

\frefformat{vario}{\fancyrefseclabelprefix}{Section~#1}
\frefformat{vario}{\fancyrefthmlabelprefix}{Theorem~#1}
\frefformat{vario}{\fancyreftbllabelprefix}{Table~#1}
\frefformat{vario}{\fancyreflemlabelprefix}{Lemma~#1}
\frefformat{vario}{\fancyrefcorlabelprefix}{Corollary~#1}
\frefformat{vario}{\fancyrefdeflabelprefix}{Definition~#1}
\frefformat{vario}{\fancyreffiglabelprefix}{Fig.~#1}
\frefformat{vario}{\fancyrefapplabelprefix}{Appendix~#1}
\frefformat{vario}{\fancyrefeqlabelprefix}{(#1)}
\frefformat{vario}{\fancyrefproplabelprefix}{Proposition~#1}
\frefformat{vario}{\fancyrefexmpllabelprefix}{Example~#1}
\frefformat{vario}{\fancyrefalglabelprefix}{Algorithm~#1}

 \newtheorem{thm}{Theorem}
 \newtheorem{cor}[thm]{Corollary}   % Turned off theorem numbering
 
 \newtheorem{defi}{Definition}
 \newtheorem{lem}[thm]{Lemma}
 
\safemath{\dictab}{[\,\dicta\,\,\dictb\,]}

\safemath{\ysig}{\bmy}
\safemath{\ysighat}{\hat{\ysig}}
\safemath{\ysigdim}{M}
\safemath{\xsig}{\bmx}
\safemath{\xsigdim}{N}
\safemath{\nx}{n_x}
\safemath{\zsig}{\bmz}
\safemath{\zsigdim}{\ysigdim}
\safemath{\rsig}{\bmr}
\safemath{\Adict}{\bA}
\safemath{\Adicttilde}{\widetilde{\Adict}}
\safemath{\Adictdim}{\outputdim\times\xsigdim}
\safemath{\avec}{\bma}
\safemath{\avectilde}{\tilde{\avec}}
\safemath{\Bdict}{\bB}
\safemath{\Bdicttilde}{\widetilde{\Bdict}}
\safemath{\Cdict}{\bC}
\safemath{\cvec}{\bmc}
\safemath{\Ddict}{\bD}
\safemath{\Ddictdim}{\ysigdim\times\xsigdim}
\safemath{\dvec}{\bmd}
\safemath{\Ddicttilde}{\widetilde{\bD}}
\safemath{\Bonb}{\bB}
\safemath{\bvec}{\bmb}
\safemath{\Bonbdim}{\ysigdim\times\ysigdim}
\safemath{\noise}{\bmn}
\safemath{\noisedim}{\ysigim}
\safemath{\err}{\bme}
\safemath{\errdim}{\ysigdim}
\safemath{\errset}{\setE}
\safemath{\nerr}{n_e}
\safemath{\delop}{\bP_\errset}
\safemath{\delopc}{\bP_{{\errset}^c}}

\safemath{\cplxi}{\imath}
\safemath{\cplxj}{\jmath}
\safemath{\dict}{\matD}
\safemath{\inputdim}{N}		% number of columns of dictionary D
\safemath{\outputdim}{M}		%number of rows of dictionary D
\safemath{\sparsity}{S}	%sparsity
\safemath{\inputdimA}{{N_a}}	%total number of elements in dictionary A
\safemath{\inputdimB}{{N_b}}	%total number of elements in dictionary B
\safemath{\elemA}{{n_a}}	%number of elements chosen from dictionary A
\safemath{\elemB}{{n_b}}	%number of elements chosen from dictionary B
\safemath{\resA}{\matR_a}	%restriction map to elements of dictionary A
\safemath{\resB}{\matR_b}	%restriction map to elements of dictionary B
\safemath{\subD}{\matS} %subdictionary
\safemath{\subA}{\matS_a} %subdictionary part of A
\safemath{\subB}{\matS_b} %subdictionary part of B
\safemath{\dicta}{\matA} 	% first subdictionary
\safemath{\dictb}{\matB} 	% second subdictionary
\safemath{\hollowS}{H}
\safemath{\hollowA}{H_a}
\safemath{\hollowB}{H_b}
\safemath{\cross}{Z}
\safemath{\coh}{\mu_d}			% coherence dictionary
\safemath{\coha}{\mu_a}			% coherence first subdictionary
\safemath{\cohb}{\mu_b}			% coherence second subdictionary
\safemath{\mubs}{\nu}	%block sub-coherence
\safemath{\cohm}{\mu_m} %mutual coherence
\safemath{\dictset}{\setD}	% set of dictionaries
\safemath{\dictsetp}{\dictset(\coh,\coha,\cohb)}	% set of dictionaries parametrized
\safemath{\dictsetgen}{\dictset_\text{gen}}
\safemath{\dictsetgenp}{\dictsetgen(\coh)}
\safemath{\dictsetonb}{\dictset_\text{onb}}
\safemath{\dictsetonbp}{\dictsetonb(\coh)}

\safemath{\leftside}{U}
\safemath{\rightsideA}{R_a}
\safemath{\rightsideB}{R_b}

\safemath{\indexS}{\setI_S} %set of indices participating in sub-dictionary S

\safemath{\na}{n_a}			% cardinality of set of linearly independent columns of first dictionary
\safemath{\nb}{n_b}			% cardinality of set of linearly independent columns of second dictionary
\safemath{\coeffa}{p_i}	%coefficients for columns of A
\safemath{\coeffb}{q_j}	%coefficients for columns of B
\safemath{\seta}{\setP}		% set of linearly independent columns of A
\safemath{\setb}{\setQ}     % set of linearly independent columns of B
\safemath{\setw}{\setW}	%set of n largest elements of w
\safemath{\setz}{\setZ}	%set of L-n largest elements of z
\safemath{\cola}{\veca}		% generic element of the dictionary A
\safemath{\colb}{\vecb}		% generic element of the dictionary B
\safemath{\cold}{\vecd}		% generic element of the dictionary D
\safemath{\inputvec}{\vecx} 	%coefficient vector (input)
\safemath{\error}{\vece}	%error vector
\safemath{\noiseout}{\vecz} 	%noisy output vector
\safemath{\inputvecel}{x}
\safemath{\inputveca}{\vecx_a}
\safemath{\inputvecb}{\vecx_b}
\safemath{\outputvec}{\vecy}	%output of Dictionary
\safemath{\lambdamin}{\lambda_{\mathrm{min}}}
\safemath{\elltwo}{\ell_2}
\safemath{\ellone}{\ell_1}
\safemath{\ellzero}{\ell_0}
\safemath{\ellinf}{\ell_\infty}
\safemath{\ellinftilde}{\ell_{\widetilde\infty}}
\safemath{\licard}{Z(\coh,\coha,\cohb)}
\safemath{\xsol}{\hat{x}}
\safemath{\xbord}{x_b}		%Solution at the border
\safemath{\xstat}{x_s}		%Solution stationary in l0 prob
\safemath{\xstatLone}{\tilde{x}_s}
\safemath{\order}{\mathcal{O}} %order notation (big O)
\safemath{\scales}{\Theta} %scales as
\safemath{\ones}{\mathbf{1}} %all ones matrix
\safemath{\zeroes}{\mathbf{0}} %all zeroes matrix
\safemath{\thlone}{\kappa(\coh,\cohb)} %treshold l1 problem
\safemath{\constoneA}{\delta} %constant in l1 theorem to save space
\safemath{\constoneB}{\epsilon} %constant in l1 theorem to save space
\safemath{\nlarge}{L}				   %num large elements
\safemath{\sumlarge}{S_\nlarge}
\safemath{\maxlarger}{P_\nlarge}	   % maximum in Gribonval and Nielsen
\safemath{\Pzero}{\textrm{P0}}	
\safemath{\Pone}{\textrm{P1}}
\safemath{\vecfir}{\vecw}			 % \vecv element of the kernel of the dictionary, \vecv=[\vecfir \vecsec]
\safemath{\vecsec}{\vecz}
\safemath{\elvecfir}{w}              % element of vecfir
\safemath{\elvecsec}{z}				 % element of vecsec
\safemath{\nlargefir}{n}
\safemath{\normout}{\gamma}
\safemath{\auxfun}{h}
\safemath{\supp}{\textrm{supp}}%support

\safemath{\indexa}{\ell}
\safemath{\indexb}{r}
\safemath{\indexc}{i}
\safemath{\indexd}{j}

\safemath{\project}{P}%projector

\usepackage{color}

\usepackage{enumitem}
\setlist[itemize]{leftmargin=*, itemsep=0.3em, topsep=0.3em} % makes itemization a bit more compact (no indents)

\allowdisplaybreaks % THIS ALLOWS EQUATIONS TO BREAK THE PAGE

\safemath{\LAMA}{\textrm{LAMA}}
\safemath{\MRT}{\textrm{MRT}}
\safemath{\betamax}{\beta^\text{max}_\setO}
\safemath{\betamaxno}{\beta^\text{max}}
\safemath{\betamin}{\beta^\text{min}_\setO}
\safemath{\betaminno}{\beta^\text{min}}

\safemath{\Nomin}{\No^\textnormal{min}(\beta)}
\safemath{\Nominnobeta}{\No^\text{min}}
\safemath{\Nomax}{\No^\textnormal{max}(\beta)}
\safemath{\Nomaxnobeta}{\No^\textnormal{max}}
\safemath{\EX}{E_\textnormal{x}}
\safemath{\EXP}{\EX^\textnormal{p}}

\safemath{\tmax}{{t_\textnormal{max}}}
\safemath{\MAP}{\textrm{MAP}}
\safemath{\IO}{\textrm{IO}}
\safemath{\JO}{\textrm{JO}}
\safemath{\Nopost}{N_{0}^\textnormal{post}}
\safemath{\MT}{U}
\safemath{\MR}{B}
\safemath{\Tran}{\textnormal{T}}
\safemath{\Herm}{\textnormal{H}}
\safemath{\row}{\textnormal{r}}
\safemath{\col}{\textnormal{c}}

\safemath{\NT}{N_\textnormal{T}}
\safemath{\DSNR}{\delta \textnormal{SNR}}
\safemath{\betaMOR}{\beta^{\star}}
\begin{document}
	
\title{Optimally-Tuned Nonparametric Linear Equalization for Massive MU-MIMO Systems
% NOPE: A Nonparametric Equalizer for massive MU-MIMO Systems
}

\author{Ramina Ghods, Charles Jeon, Gulnar Mirza, Arian Maleki, and Christoph Studer\thanks{R.~Ghods, C.~Jeon, G.~Mirza, and C.~Studer are with the School of ECE at Cornell University, Ithaca, NY; e-mails: {rg548@cornell.edu}, {jeon@csl.cornell.edu}, {gzm3@cornell.edu}, and {studer@cornell.edu}.}\thanks{A. Maleki is with Department of Statistics at Columbia University, New York City, NY; e-mail: {arian@stat.columbia.edu}.}\thanks{The work of RG, CJ,  and CS was supported in part by Xilinx, Inc.~and by the US National Science Foundation~(NSF) under grants ECCS-1408006,  CCF-1420328, and CAREER CCF-1652065.}
}

\maketitle

\begin{abstract}
This paper deals with linear equalization in massive multi-user multiple-input multiple-output (MU-MIMO) wireless systems. 
We first provide simple conditions on the antenna configuration for which the well-known linear minimum mean-square error (L-MMSE)  equalizer provides near-optimal spectral efficiency, and we analyze its performance in the presence of parameter mismatches in the signal and/or noise powers.
We then propose a novel, optimally-tuned NOnParametric Equalizer (NOPE) for massive MU-MIMO systems, which avoids knowledge of the transmit signal and noise powers altogether.
We show that NOPE achieves the same performance as that of the L-MMSE equalizer in the large-antenna limit, and we demonstrate its efficacy in realistic, finite-dimensional systems.
From a practical perspective, NOPE is computationally efficient and avoids dedicated training that is typically required for parameter estimation.
%(short for \underline{no}ne \underline{p}arametric \underline{e}qualizer).
%
\end{abstract}

% This paper deals with linear equalization in massive multi-user multiple-input multiple-output (MU-MIMO) wireless systems. We first provide simple conditions on the antenna configuration for which the well-known linear minimum mean-square error (L-MMSE)  equalizer provides near-optimal spectral efficiency, and we analyze its performance in the presence of parameter mismatches in the signal and/or noise powers. We then propose a novel, optimally-tuned NOnParametric Equalizer (NOPE) for massive MU-MIMO systems, which avoids knowledge of the transmit signal and noise powers altogether. We show that NOPE achieves the same performance as that of the L-MMSE equalizer in the large-antenna limit, and we demonstrate its efficacy in realistic, finite-dimensional systems. From a practical perspective, NOPE is computationally efficient and avoids dedicated training that is typically required for parameter estimation.

\section{Introduction}
Massive (or large-scale) multi-user multiple-input multiple-output (MU-MIMO) will be among the key technologies for fifth-generation~(5G) wireless systems as it provides (often significantly) higher spectral efficiency than traditional, small-scale MIMO~\cite{LETM2014,ABCHLAZ2014}. 
Data detection at the infrastructure base-stations (BSs) in such massive MU-MIMO systems is among the most critical components from a spectral efficiency and computational complexity  perspective. 
In particular, since optimal data detection is known to be an NP-hard problem~\cite{V1998}, a na\"ive exhaustive search over all possible transmit signals  %scales exponentially in the number of user antennas $M_T$. Hence, optimal data detection 
would result in prohibitive computational complexity for such large-dimensional systems. Hence, alternative algorithms that achieve high spectral efficiency at low complexity must  be deployed in practice. 
In addition, practical massive MU-MIMO systems suffer---as do traditional MIMO systems---from real-world hardware impairments and model mismatches (for example amplifier nonlinearities, phase noise, quantization artifacts, channel-estimation errors, etc.). Such system nonidealities are known to substantially reduce the performance of optimal data-detection algorithms unless one explicitly models these impairments and estimates the associated parameters  \cite{Studer_Tx_OFDM}. 

\subsection{Contributions}
In this paper, we address the above challenges by developing a novel, \underline{no}n\underline{p}arametric \underline{e}qualizer (NOPE) for massive MU-MIMO systems that requires low complexity and is robust to system impairments and model mismatches. Our key contributions can be  summarized as follows:
%	In \fref{sec:L-equalization}, we provide a thorough analysis of L-MMSE equalization in massive MIMO systems. In particular,  
\begin{itemize}
\item We present simple conditions for which linear minimum mean-squared error (L-MMSE), zero-forcing (ZF), and maximum ratio combining (MRC)-based equalizers provide near-optimal performance in massive MU-MIMO systems.
\item We analyze the impact of parameter mismatches on  L-MMSE equalization by extending the results by Tse and Hanly in~\cite{TH1999}.
\item We propose a novel, computationally efficient, and nonparametric algorithm called  NOPE that does not require any knowledge of the signal and noise powers.
\item We prove,  in the large-antenna limit, that NOPE achieves the same performance as that of the L-MMSE equalizer, which requires knowledge of the signal and noise powers.
\item We demonstrate that NOPE achieves the performance of the L-MMSE equalizer in realistic, finite-dimensional systems.
\end{itemize}
\subsection{Relevant Prior Art}
L-MMSE estimation is used in a large number of communication applications for estimation, detection, and equalization~\cite{madhow1994mmse,xie1990family,VS1999,kumar2009asymptotic,TH1999,SV2001}. The low complexity (besides the inversion of a potentially large matrix) and acceptable performance in many situations are responsible for the widespread use of L-MMSE estimation in practical  transceiver designs.
For massive MU-MIMO systems, it was shown in~\cite{HBD11} that  L-MMSE equalization enables (often significantly) higher achievable rates than ZF or MRC-based equalizers. To complement these results, we provide conditions on the antenna configuration for which L-MMSE, ZF, and MRC approach the fundamental performance limits. 

One of the downsides of L-MMSE equalization is that it requires accurate estimates of the signal and noise powers. While Al-Dhahir and Cioffi  in  \cite{al1997mismatched} analyzed the effect of parameter mismatches to MMSE-based decision-feedback equalizers, we extend the results by Tse and Hanly in \cite{TH1999} and provide  large-antenna limit expressions for the output signal-to-inference ratio (SIR) of L-MMSE equalization in the presence of parameter mismatches. 

To mitigate the impact of parameter mismatches, linear and adaptive methods have been proposed for randomly spread code-division multiple-access (CDMA) systems in~\cite{honig1995blind,wang1998blind}. Inspired by such results, we propose an optimally-tuned, nonparametric linear equalization algorithm for massive MU-MIMO systems. Our algorithm does not require any knowledge of the signal and/or noise powers, achieves the performance of the L-MMSE equalizer in the large-antenna limit, and is computationally efficient as it avoids a costly matrix inversion.

% Robust to impairments, relatively low complexity ; relatively good performance, excellent performance in massive mimo. 
%Linear MMSE is a well-known estimation technique old and awesome; most used method in a large number of communication applications for estimation and equalization based detection. Randomly spread CDMA, MIMO, etc. Robust to impairments, relatively low complexity (besides inverting a potentially large matrix); relatively good performance, excellent performance in massive mimo. 
%
%
\subsection{Notation}
Lowercase and uppercase boldface letters designate column vectors and matrices, respectively. For a matrix $\bA$, we define its conjugate transpose as $\bA^\Herm$. The entry on the $k$-th row and $\ell$-th column is $A_{k,\ell}$, 
%and the $k$-th entry of a vector $\veca$ is~$a_k$. The $M\times M$ identity matrix is denoted by $\bI_M$ and the $M\times N$ all-zeros matrix by $\mathbf{0}_{M\times N}$. 
We define $\left\langle \bma \right\rangle = \frac{1}{N}\sum_{k=1}^N a_k$. Multivariate complex-valued Gaussian probability density functions (PDFs)  with mean vector~$\bmm$ and covariance matrix~$\bK$ are denoted by $\setC\setN(\bmm,\bK)$. Expectation and variance with respect to the PDF of the random variable $X$ is $\Exop_X\!\left[\cdot\right]$ 
and $\Varop_X\!\left[ \cdot\right] $, respectively. 
%We use $\mathbb{P}(X=x)$ to denote the probability of the RV $X$ being~$x$.
%
%
%
%\section{System Model and Linear Equalization }
%
%\subsection{System Model}
%
%
%Also put the key assumptions in here.
%
%

\section{Linear MMSE Equalization}\label{sec:L-equalization}

We start by introducing the system model and reviewing the basics of parametric L-MMSE equalization with perfect knowledge of the signal and noise powers. 
 We then develop a computationally efficient algorithm that relies on approximate message passing (AMP) and we  analyze its performance. 

\subsection{System Model}
We consider the following standard input-output relation of a massive MU-MIMO uplink system: $\bmy=\bH \bmx+\bmn$.
Here, the vector $\vecy\in\complexset^\MR$ corresponds to the received signal ($\MR$ denotes the number of BS antennas), the matrix $\bH\in\complexset^{\MR\times \MT}$ represents the uplink channel ($\MT$ denotes the number of user antennas), the transmit signal vector is $\vecx\in\complexset^\MT$, and the vector $\bmn\in\complexset^\MR$ models receive noise where all its entries are assumed to be i.i.d.\ circularly-symmetric complex Gaussian with variance~$\No$ per complex entry. 
We assume that the transmit signal vector~$\bmx$ has i.i.d entries, i.e., $p(\vecx)=\prod_{u=1}^{\MT}p(x_u)$, with zero mean and signal variance $\EX=\Ex{}{|x_u|^2}$, $\forall u$. We define the \emph{antenna ratio} as $\beta=\MT/\MR$, and the following quantities:
\begin{defi}
The \emph{large-antenna limit} is defined by fixing the antenna ratio $\beta=\MT/\MR$ and letting $\MT\to\infty$.
\end{defi}
\begin{defi}
A matrix $\bH$ has  \emph{uniform channel gains} if the entries of $\bH$ are i.i.d.\ circularly-symmetric complex Gaussian with variance $1/\MR$ per complex entry, i.e., $H_{b,u}\sim\setC\setN(0,1/B)$. 
\end{defi}

\subsection{Basics of L-MMSE Equalization}\label{sec:L-MMSE}
%
%Here, we need a kick-ass tutorial of linear equalization with all the old standard references. The standard form of MMSE and what the underlying assumptions are (Gaussian priors yield exactly the MMSE form). Show the explicit inverse form; which is well known and used widely in practice. Show special cases: ZF and MRC. 
%
%l-MMSE equalization is a linear filter that minimizes the mean square error $\textit{MSE} = \Ex{\,\bms,\bH}{\|\hat\bms-\bms_0\|^2}$. 
%\cs{maybe assume iid signals and noise; then the derivation simplifies a lot.} 
We start with the following well-known facts. The equalization matrix $\bW\in\complexset^{\MT\times \MR}$ of the L-MMSE equalizer output $\hat\bmx=\bW\bmy$ that minimizes the $\textit{MSE} = \Ex{\,\bmx,\bmn}{\|\hat\bmx-\bmx\|^2}$ is given by  
%\begin{align}
%\bW = \textstyle  \bH^H \Big(\bH\bH^H+\frac{\No}{\EX}\bI_\MR\Big)^{-1}.
%\end{align}
%where $\bK_\bms=\Ex{}{\bms\bms^H}$ and $\bK_\bmn=\Ex{}{\bmn\bmn^H}$. In non-degenerate cases where, where $\bK_\bms$ and $\bK_\bmn$ are positive definite (i.e. invertible), we have
%%%%%%%%%%% equivalent form
%This can also be written in the following equivalent form:
%\begin{align}\label{eq:LMMSE}
$\bW = (\bH^\Herm \bH+{\No }/{\EX})^{-1} \bH^\Herm$.
%\end{align}
%which turns out to require a matrix inversion of dimension $\MT\times\MT$ only. 
%%%%%%%%%%%%%
At high signal-to-noise-ratio (SNR), i.e.,  for ${\No}/{\EX} \to 0$, the L-MMSE equalizer implements zero-forcing (ZF) equalization with output $\hat\bmx= (\bH^\Herm\bH)^{-1}\bH^\Herm \bmy$; 
at low SNR, the L-MMSE equalizer implements MRC with output $\hat{\bmx}=\bH^\Herm \bmy$.
We note that L-MMSE is the optimal (linear and nonlinear) equalizer if the signal and noise are independent and both i.i.d. circularly symmetric complex Gaussian vectors. 

\subsection{L-MMSE Equalization via AMP}\label{sec:MMSE-AMP}
% 
%\cs{introduce the algorithm with mismatched SE; mismatch also in terms of No and Nopost.}
L-MMSE equalization as summarized above, can be implemented using the mismatched complex bayesian AMP (mcB-AMP) algorithm proposed in \cite{JMS2016}, which is a mismatched version of the AMP algorithm put forward in \cite{donoho2011design}. 
%
%Assume the wireless model \fref{eq:sysmodel} 
In particular, we use mcB-AMP with the following Gaussian mismatched signal prior distribution: $\tilde{p}(\vecx)=\prod_{i=1}^{N}\tilde{p}(x_i)$ with $\tilde{p}(x_i) \sim \setC\setN(0,\EX)$. By assuming that the detector knows the true transmit signal prior distribution $X_0\sim p(x_0)$ (e.g., two Dirac delta functions for BPSK), the \emph{parametric} L-MMSE algorithm, which we will refer to as MMSE-AMP, is as follows:
\newtheorem{alg}{Algorithm} 
\begin{alg} Initialize $t=1$, $x^1_\ell = \Exop_{X_0}[X_0]$, $\ell=1,\ldots,\MT$, and $\bmr^1 = \bmy - \bH\bmx^1$. Then, for every iteration $t=1,2,\ldots,\tmax$ compute the output $\bmz^t$ via the following steps:\label{alg:MMSE-AMP}
   \begin{align}
   \nonumber
   \tilde\sigma^2_{t} &= \textstyle \frac{1}{\MR}\vecnorm{\bmr^{t}}_2^2\\
   \label{eq:tau_opt}
   \tau^{t} &= \textstyle \argmin_{\tau\geq0} \,\Psi(\tilde\sigma_t^2,\tau)\\
   \nonumber
   \bmz^t&=\bmx^{t} + \bH^\Herm \bmr^{t}\\
   \label{eq:xhat_LMMSE}
   \bmx^{t+1} &= \mathsf{F}^\text{mm}\!\left(\bmz^t,\tau^t\right)\\
   \label{eq:residual}
   \bmr^{t+1} &= \bmy - \bH\bmx^{t+1} + \beta \bmr^t \!\left\langle\mathsf{F'}^\text{mm}(\bmz^t,\tau^t)
   \right\rangle\!,
   \end{align}
%   which is carried out for $\tmax$ iterations $t=1,\ldots,\tmax$.
%   % 
%   The algorithm is initialized by $x^1_\ell = \Exop_{X_0}[X_0]$ for all $\ell=1,\ldots,\MT$, $\bmr^1 = \bmy - \bH\bmx^1$. 
Here, $\mathsf{F}^\text{mm}(x_\ell,\tau)=\frac{\EX}{\EX+\tau}x_\ell$ is the posterior mean function, $\mathsf{F'}^\text{mm}(x_\ell,\tau)$ is its derivative in the first argument, and both functions operate element-wise on vectors. Furthermore,
\begin{align*}
\textstyle \Psi(\tilde\sigma_t^2,\tau)=\Exop_{X_0,Z}\!\left[\abs{\mathsf{F}^\text{mm}(X_0+\tilde\sigma_t Z,\tau)-X_0}^2\right]\! =\frac{\tau^2 \EX+\sigma_t^2\EX^2}{(\EX+\tau)^2}
\end{align*}
is the MSE function in which the expectation is taken with respect to the true signal prior $X_0\sim p(x_0)$ and $Z \sim \setC \setN(0,1)$. 
\end{alg}

We note that \fref{alg:MMSE-AMP} avoids the computation of matrix inverses, which often dominate the computational complexity of L-MMSE equalizers in small and large MIMO systems \cite{studer2011asic}.  

%\begin{algorithm}
%	\caption{MM-LAMA}
%	\label{alg:MM-LAMA}
%	\begin{algorithmic}
%		\STATE {\bf inputs:} $\bH\in\complexset^{\MR\times \MT}$ and $\bmy\in\complexset^{\MR}$
%		% 
%		\STATE {\bf initialize:} $t=1$, $x^1_\ell = \Exop_{X_0}[X_0]$ for all $\ell=1,\ldots,\MT$, $\bmr^1 = \bmy - \bH\bmx^1$
%		\FOR {$t\leq \tmax$}
%		\STATE $\tilde\sigma^2_{t} = \textstyle \frac{1}{\MR}\vecnorm{\bmr^{t}}_2^2$
%		\STATE $\tau^{t} = \argmin_{\tau\geq0} \,\Psi(\tilde\sigma_t^2,\tau)\!$
%		\STATE $\bmz^t=\bmx^{t} + \bH^\Herm \bmr^{t}$
%		\STATE $\bmx^{t+1}= \mathsf{F}^\text{mm}\!\left(\bmz^t,\tau^t\right)\!$
%		\STATE $\bmr^{t+1} = \bmy - \bH\bmx^{t+1} + \beta \bmr^t \!\left\langle\mathsf{F'}^\text{mm}(\bmz^t,\tau^t)\right\rangle\!$
%		\STATE $t=t+1$
%		\ENDFOR
%		\STATE {\bf output:} Gaussian output ${\bmz}^t$, where each user has variance $\|\bmr^t\|^2/\MR $
%		% 
%	\end{algorithmic}
%\end{algorithm}
%
%
\subsection{Asymptotic Performance Analysis}\label{sec:LMMSE-SE}
One of the key properties of AMP-based algorithms is that their SIR performance can be analyzed  in the large-antenna limit using the state evolution (SE) framework. 
We require the following result; the proof follows from \cite[Eq.~12]{JMS2016}: 
\begin{lem} \label{lem:SE}
Fix $\beta$ and let $\bH$ have uniform channel gains. Then,  in the large-antenna limit, the output $\bmz^t$ of \fref{alg:MMSE-AMP} can be modeled as $\bmz^t=\bmx+\bmw^t$ with $\bmw^t \sim \setC \setN(0,\sigma_t^2 \bI_{\MT})$, where
%the MIMO system is effectively decoupled into a set of MT parallel and independent additive white Gaussian noise (AWGN) channels. For the MMSE-AMP algorithm above, 
the equivalent noise variance $\sigma_t^2$ in iteration $t$ is given by the following SE equation:
%\begin{align}\label{eq:LMMSE-SE}
$\sigma_{t+1}^2=\No+\beta \frac{\EX}{\EX+\sigma_t^2}\sigma_t^2$.
%\end{align}
\end{lem}
For $t \to \infty$, the SE equation coincides with the asymptotic SIR expression of the L-MMSE equalizer given by Tse and Hanly in \cite[Thm.~3.1]{TH1999} with $\textit{SIR}=1/\sigma^2$. Hence. we have:

\begin{cor} \label{cor:mmseopt}
In the large-antenna limit, MMSE-AMP achieves the same SIR performance as the L-MMSE equalizer. 
\end{cor}
Since in the large-antenna limit and for uniform channel gains AMP-based equalization decouples the MIMO system into parallel and independent AWGN channels with variance~$\sigma^2_t$ (see \cite[Sec.~6]{andreaGMCS} for the details of this decoupling property), the per-user achievable rate of the L-MMSE equalizer is given by
%\begin{align}\label{eq:C_MMSE}
$C_\text{L-MMSE}=\log_2\big(1+{\EX}/{\sigma^2}\big)$ [bits/user/channel use],
%\end{align}
where~$\sigma^2$ is the fixed-point to the SE equation in \fref{lem:SE}.
%Here, we will recover classical results from literature; mention how it can be used to predict the performance of QPSK etc; also talk about the beta min max issue; issue etc; Introduce the mismatched state evolution framework; then use it to derive the equations. 
%

\subsection{When Does L-MMSE Achieve Near-Optimal Performance?}\label{sec:optimality}
%
%In \cite{JGMS2015conf,JGMS2015} we developed and analyzed a novel data detection algorithm suitable for massive MU-MIMO systems using approximate message passing (AMP)  called LAMA. We showed how LAMA is optimal in the large-antenna limit and achieves near-optimal performance in practical (finite-dimensional) systems at low complexity. 
%
As shown in \cite{VS1999,kumar2009asymptotic,hoydis2013massive} for massive MU-MIMO systems with significantly more BS antennas than users (i.e., for small values of~$\beta$), linear equalizers, such as MRC, ZF, and L-MMSE achieve near-optimal performance. More specifically, reference~\cite{hoydis2013massive} derives conditions for which ZF and MRC approach the performance of L-MMSE equalization. We now  provide conditions on the antenna ratio~$\beta$ for which linear equalizers approach \emph{optimal} performance. 
We use the fact that,  in the large-antenna limit, the individually-optimal (IO) posterior mean estimator (PME) decouples the MIMO system into~$U$ parallel and independent AWGN channels $\bmz=\bmx+\bmw$ with $\bmw \sim \setC \setN(0,\sigma^2 \bI_{\MT})$, where the equivalent noise variance satisfies $\sigma^2\geq\No$~\cite{GV2005}. 
Hence, the fundamental performance limit is given by the channel capacity of an AWGN channel $C_\text{AWGN}=\log_2\!\big(1+{\EX}/{\No}\big).$
We therefore characterize the performance of linear equalizers as follows:
\begin{defi}\label{def:dSNR}
Assume that the IO-PME and a linear equalizer achieve the same rate $R$ at SNR $\EX/\No$ and $\EX/\hat{\No}$, respectively. We define the \emph{SNR loss} as
%\begin{align*}
%\DSNR &= 10 \log_{10}\left(\frac{\No}{\hat{\No}}\right).
%\begin{align*}
$\DSNR = {\No}/\hat{\No}$,
%\end{align*}
which satisfies $1\leq\DSNR$ and characterizes the excess SNR required  by a linear equalizer to achieve optimal performance.
% a given rate $R$.
% rate $C_\text{AWGN}$.	
\end{defi}

For MRC, ZF, and L-MMSE equalization, we can establish the following result with proof given in~\fref{app:MOR}.
\begin{lem}\label{lem:MOR}
Assume the large-antenna limit and let $\bH$ have uniform channel gains. For a fixed rate~$R$, the SNR loss of MRC, ZF, and L-MMSE will be no larger than $\DSNR$ if $\beta \leq \betaMOR(\DSNR,R)$, where $\betaMOR(\DSNR,R)$ is the \emph{maximum optimal antenna ratio (MOAR)} given by
\begin{align*}
%\text{MRC:}& & \betaMOR_R(\DSNR^*)&=\frac{1-10^{-\DSNR^*/10}}{2^R-1} \\
%\text{ZF:}  & & \betaMOR_R(\DSNR^*)&=1-10^{-\DSNR^*/10} \\
%\text{L-MMSE:}& & \betaMOR_R(\DSNR^*)&=\left(1-10^{\frac{-\DSNR^*}{10}} \right)\frac{2^R}{2^R-1}
\text{MRC:}& & \betaMOR(\DSNR,R)&= \textstyle \left(1-\DSNR^{-1}\right)\! \frac{1}{2^R-1} \\
\text{ZF:}  & & \betaMOR(\DSNR,R)&= \textstyle1-\DSNR^{-1} \\
\text{L-MMSE:}& & \betaMOR(\DSNR,R)&= \textstyle \left(1-\DSNR^{-1} \right)\!\frac{2^R}{2^R-1}.
\end{align*}
\end{lem}

\setlength{\textfloatsep}{6pt}% Remove \textfloatsep
\begin{figure}
\centering
\includegraphics[width=0.8\columnwidth]{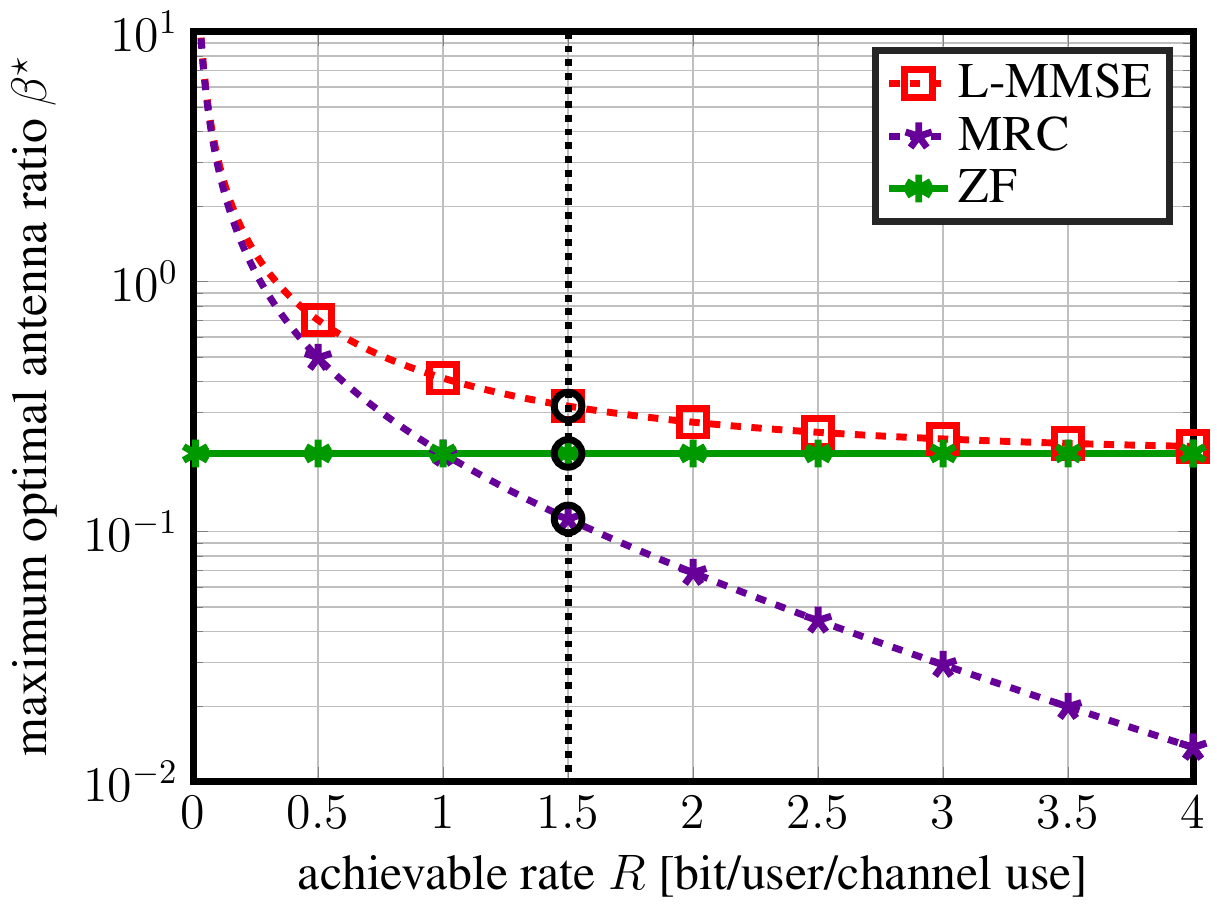}\label{fig:MOR}
\vspace{-0.2cm}
\caption{Comparison of the maximum optimal antenna ratio (MOAR) $\betaMOR$ for MRC, ZF, and L-MMSE equalization to operate within $\DSNR=1$\,dB of the fundamental performance limit. For all achievable rates, L-MMSE supports the largest antenna ratio $\beta=\MT/\MR$ to operate within $1$\,dB of optimal performance. }
\label{fig:MOR}
\end{figure}
%
%Here, we should compare MMSE to LAMA and give guidelines for what beta linear is good enough; may depend on the modulation scheme and on No. For large No, we should see that linear equalization is good enough. For small No, it will depend on beta. Maybe there's something like No*beta that tells us when it is good enough. A simulation would be easy---but theory would be super much cooler.  
%
\fref{fig:MOR} illustrates the MOAR for MRC, ZF, and L-MMSE equalization as a function of their achievable rates for a fixed SNR loss of $\DSNR=1$\,dB.  \fref{lem:MOR} identifies the maximum antenna ratio $\beta$ for which these linear equalizers are able to operate within $1$\,dB SNR of the fundamental performance limit.
We see that L-MMSE enables the largest user-to-BS antenna ratios $\beta$ among the considered equalizers for all rates.
Consider, for example,  the dotted black vertical line at $1.5$\,bit/user/channel use. We see that L-MMSE exhibits an SNR loss of less than 1\,dB for any antenna ratio below 0.3;  MRC and ZF require antenna ratios of 0.2 and 0.1, respectively. 
%
%We furthermore see that MRC performs as well as L-MMSE for very low rates $R$ (or low values of SNR);  for high rates, ZF performs as well as L-MMSE equalization. 

%
%\rg{commented $\betaMOR$ for mismatch since we are not putting it in the MOR plot.}
%%%%%%%%%%%%%%%%%%%%%%%%%%%%%%%%%%%%%%%%%%%%%%%%%%%%%%%%%%%%%%%%%%%%%%%%%%%%%%%%
%\begin{lem}\label{lem:MMM_MOR}
%Hence, for a given capacity $C$, the maximum antenna-ratio $\betaMOR$ available for optimal performance in linear MMSE with parameter mismatch is given by the solution to the following coupled equations:
%\begin{align*}
%1&= 10^{\frac{-\Delta \text{SNR}}{10}} + \betaMOR \frac{\theta^4 (2^C-1)+{\EXP}}{(\EXP+\theta^2)^2},\\
%\theta^2&=\frac{10^{\frac{-\Delta \text{SNR}}{10}} }{2^C-1}\EX+\betaMOR \frac{\theta^2 \EXP}{\theta^2+\EXP}
%\end{align*}
%and $\Delta \text{SNR}$ is the SNR error gap in dB. 
%\end{lem}
%Here, we skip the derivation for \fref{lem:MMM_MOR} as it follows directly that of \fref{thm:MOR}.
%%%%%%%%%%%%%%%%%%%%%%%%%%%%%%%%%%%%%%%%%%%%%%%%%%%%%%%%%%%%%%%%%%%%%%%%%%%%%%%%
%
%%
%
%
%
%
\section{L-MMSE with Parameter Mismatch}\label{sec:MM}
Although L-MMSE significantly outperforms MRC and ZF equalization, it requires knowledge of the signal and noise powers. 
We now analyze the impact of a signal power parameter mismatch on the performance of the L-MMSE equalizer. 

\subsection{Mismatch Analysis of AMP-based L-MMSE equalizer}
Analogously to \cite[Sec.~\uppercase\expandafter{\romannumeral 3}-A]{JMS2016}, the mcB-AMP algorithm with a mismatched Gaussian  prior $p(x_u) = \setC \setN(0,\EXP)$, $\forall u$, achieves the performance of a L-MMSE equalizer with \emph{mismatched signal power} defined by $\EXP$. The following lemma provides the SE equation for this mismatched L-MMSE algorithm;  the proof can be established from the results of the coupled SE equations in~\cite[Thm.~1]{JMS2016}:
\begin{lem}\label{lem:MMM_SE}
Fix $\beta$ and let $\bH$ have uniform channel gains. Then,  in the large-antenna limit, the performance of an optimally tuned AMP-based L-MMSE estimator with a mismatched signal power $\EXP$ is given by the following coupled SE equations:
	\begin{align*}
	\sigma_{t+1}^2= \textstyle \No+\beta \frac{\theta_t^4 \EX+{\EXP}^2\sigma_t^2}{(\EXP+\theta_t^2)^2}  \quad \text{and} \quad
	\theta_{t+1}^2=\textstyle \No+\beta \frac{\EXP \theta_t^2}{\EXP+\theta_t^2}.
	\end{align*}
\end{lem}
%Similar to the previous sections, by setting $\sigma^2_\text{MM}=\sigma_{t}^2$ and letting $t \to \infty$ in the SE equation in \fref{lem:MMM_SE}, the capacity of linear MMSE equalizer with mismatched signal power corresponds to $C=\log_2(1+\frac{\EX}{\sigma^2_\text{MM}})$.
%
We note that  for $t \to \infty$ and in the large-antenna limit, the result from \fref{lem:MMM_SE} empirically matches the performance of the standard L-MMSE estimator in \fref{sec:L-MMSE} with $\No/\EXP$.
For no parameter mismatch, i.e., for $\EXP=\EX$, we have:
\begin{cor}
Let $\EX=\EXP$. Then, the SE equations in \fref{lem:MMM_SE} for $t \to \infty$ coincide with the Tse-Hanly equation in \cite[Thm.~3.1]{TH1999} for the L-MMSE equalizer with $\text{SIR}=1/\sigma^2$.
\end{cor}

\subsection{Numerical Analysis of Signal Power Mismatch}
\begin{figure}
\centering
\includegraphics[width=0.8\columnwidth]{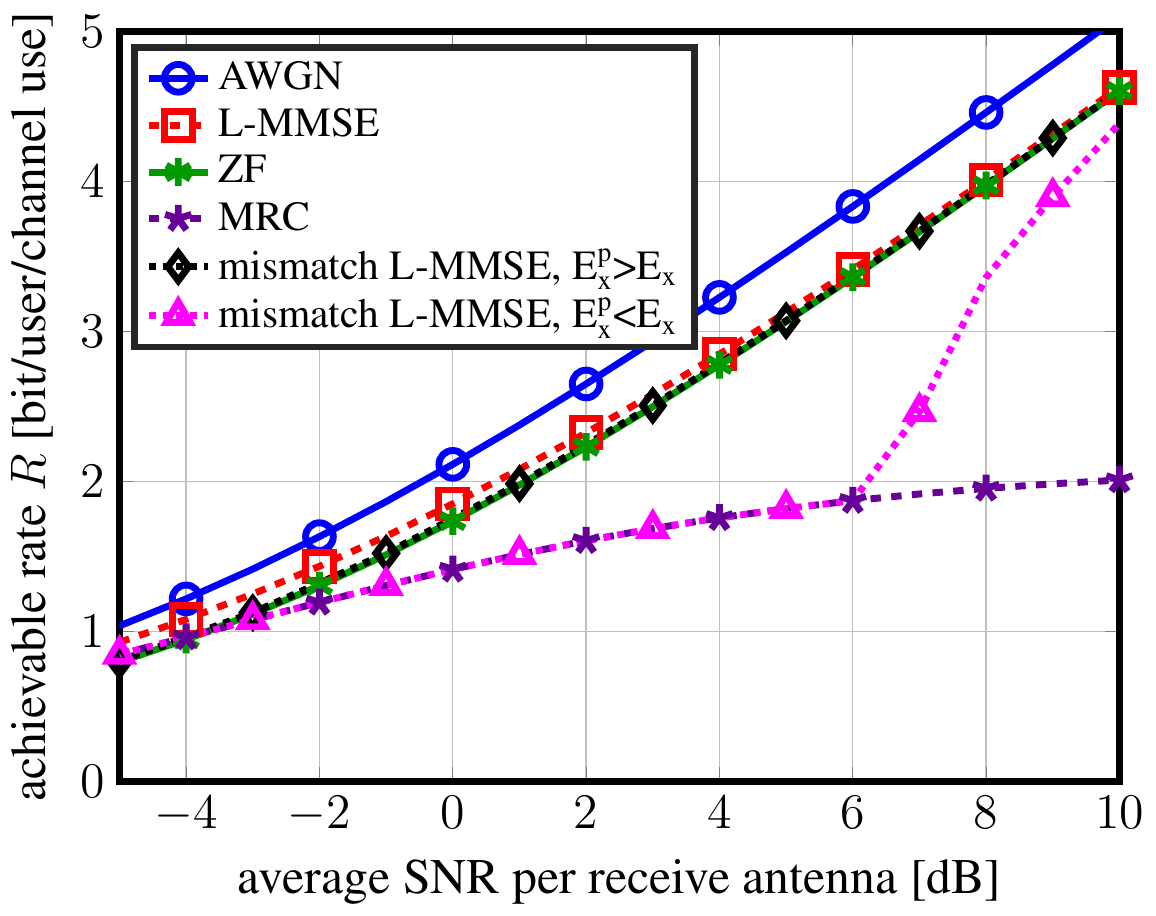}\label{fig:capacity}
\vspace{-0.22cm}
\caption{Achievable rates for different equalizers with $\beta=0.3$. Overestimating ($\EXP>\EX$) and underestimating $\EXP<\EX$) the signal power in the L-MMSE equalizer entails significant losses in terms of the achievable rates.}
\label{fig:capacity}
\end{figure}
\fref{fig:capacity} illustrates the impact of parameter mismatch on the  achievable rate of L-MMSE equalization in a MIMO system with an antenna ratio of $\beta=0.3$. The dotted black and magenta curves show the achievable rate of the mismatched L-MMSE equalizer with overestimated and underestimated signal powers. The mismatched signal power $\EXP$ is set to the 90-percentile of a signal-power estimator with two training data symbols. 
In words, for a given SNR, the mismatched L-MMSE equalizer will have an achievable rate between that of the exact L-MMSE equalizer (red curve) and the minimum of the mismatched versions (black and magenta dotted curves) in $90\%$ of the transmissions. Clearly, the mismatched L-MMSE equalizer may experience a significant rate loss.
We also see that underestimating the signal power ($\EXP<\EX$) results in a performance close to that of MRC; overestimating  ($\EXP>\EX$) results in a performance close to that of the ZF equalizer.

\section{NOPE: Nonparametric Equalizer}\label{sec:NOPE}
To cope with the detrimental effects of parameter mismatches in the L-MMSE equalizer, we now present our NOPE algorithm, which does not require knowledge of the signal or noise powers. 
\subsection{The NOPE Algorithm} \label{sec:NOPE1}
MMSE-AMP, as in \fref{alg:MMSE-AMP}, requires knowledge of the true transmit signal prior distribution $p(\bmx)$ and the signal power~$\EX$. 
To design a nonparametric version that does not need this information, we need to solve the following issues:
\begin{itemize}
\item NOPE will not have any knowledge of the true signal prior. Hence, to tune the threshold parameter $\tau^{t}$ in \fref{eq:tau_opt}, we need a way to estimate the  MSE function $\Psi(\tilde\sigma_t^2,\tau^t)$.
\item The signal power $\EX$ must be tuned in each algorithm iteration $t$ to achieve optimal performance
\end{itemize}
Reference \cite{mousavi2015consistent} develops a nonparametric approach to tune the threshold parameter of AMP-based sparse signal recovery using Stein's unbiased risk estimate (SURE)\cite{stein1981estimation}. Inspired by this approach, we will tune both the (unknown) signal power~$\EX$ and the parameter $\tau^t$ using SURE. 
We start by defining the parameter $\gamma^t={\EX}/{\tau^t}$ and rewrite the functions $\mathsf{F}^\text{mm}(x_\ell,\gamma^t)=\frac{\gamma^t}{\gamma^t+1}x_\ell$ and $\mathsf{F'}^\text{mm}(x_\ell,\gamma^t)=\frac{\gamma^t}{\gamma^t+1}$ in \fref{alg:MMSE-AMP}. As a consequence of this parameter change, we only need to estimate a single parameter per iteration, namely~$\gamma^t$.
   
Optimal tuning within NOPE can be achieved if the parameters $\{\gamma^1,\dots,\gamma^{\tmax}\}$ are tuned so that the MSE function $\Psi$ is minimized at iteration $\tmax$. As shown in \cite[Thm.~3]{JMS2016} a joint optimization over all parameters is not required---instead optimal tuning can be achieved by tuning each parameter~$\gamma^t$ separately at iteration $t$ starting from $t=1$ to $\tmax$.

To estimate the MSE function~$\Psi$ without knowledge of the true signal prior, we use the SURE function, which is  given by
\begin{align}
   \nonumber \hat{\Psi}(\tilde\sigma_t^2,\gamma^t)=&\,\textstyle
   \frac{1}{\MT}\|\mathsf{F}^\text{mm}(\bmz^{t},\gamma^t)-\bmz^{t}\|^2\\
   \nonumber
   & \textstyle +\tilde\sigma_t^2+2\tilde\sigma_t^2\left\langle\mathsf{F'}^\text{mm}(\bmz^{t},\gamma^t)-1\right\rangle\\
   =& \, \textstyle\, \tilde\sigma_t^2\frac{\gamma^t-1}{\gamma^t+1}+\frac{\|\bmz^t\|_2^2}{\MT(\gamma^t+1)^2}.\label{eq:Psi_hat}
\end{align}
The minimum of the SURE function is achieved for $\gamma_\text{min}^{t}\!=\!{\|\bmz^t\|_2^2}/(\MT\tilde{\sigma}_t^2)-1$. Hence, we replace the tuning stage in \fref{eq:tau_opt} by $\gamma_\text{min}^{t}$ to arrive at the following algorithm we call NOPE:

%\begin{algorithm}
%	\caption{NOPE}
%	\label{alg:NOPE1}
%	\begin{algorithmic}
%		\STATE {\bf inputs:} $\bH\in\complexset^{\MR\times \MT}$ and $\bmy\in\complexset^{\MR}$
%		\STATE {\bf precompute:} $\hat{d}_\ell^2 = \sum_{j=1}^\MR \abs{H_{j,\ell}}^2$, $\ell=1,\ldots,\MT$, $\hat{d}_\text{avg} = \frac{1}{\MT}\sum_{\ell=1}^\MT\hat{d}_\ell^2$
%		% 
%		\STATE {\bf initialize:} $t=1$, $\bmr^t=\bmy$, $\bmx^t=\bf0$
%		\FOR {$t\leq \tmax$}
%		\STATE $\bmz^t=\bmx^{t} + \bH^\Herm \bmr^{t}$
%		\STATE $\gamma^{t} = \frac{1}{\beta}\frac{\|\bmz^t\|_2^2}{\|\bmr^t\|_2^2} - 1$
%		\STATE $\bmx^{t+1} = \frac{\gamma^t}{\gamma^t+1}\bmz^t$
%		\STATE $\bmr^{t+1} = \bmy - \bH\bmx^{t+1} + \beta \bmr^t \!\frac{\gamma^t}{\gamma^t+1}\!$
%		\STATE $t=t+1$
%		\ENDFOR
%		\STATE {\bf output:} Gaussian output ${\bmz}^t$, where each user has variance $\|\bmr^t\|^2/\MR $
%% 
%\end{algorithmic}
%\end{algorithm}
%\newtheorem{alg}{Algorithm}
\begin{alg}
Initialize $t=1$, $\bmr^1=\bmy$, and $\bmx^1=\bf0$. Then, for every iteration $t=1,2,\ldots,\tmax$ compute the following steps:\label{alg:NOPE1}
\begin{align*}
   \bmz^t&=\bmx^{t} + \bH^\Herm \bmr^{t}\\
  \gamma^{t} &= \textstyle \frac{1}{\beta}\frac{\|\bmz^t\|_2^2}{\|\bmr^t\|_2^2} - 1\\
 \bmx^{t+1} &= \textstyle \frac{\gamma^t}{\gamma^t+1}\bmz^t\\
 \bmr^{t+1} &=\textstyle  \bmy - \bH\bmx^{t+1} + \beta \bmr^t \!\frac{\gamma^t}{\gamma^t+1}.
\end{align*}
\end{alg}
%   \begin{align*}
%   \bmz^t&=\bmx^{t} + \bH^\Herm \bmr^{t}\\
%   \gamma^{t} &= \frac{1}{\beta}\frac{\|\bmz^t\|_2^2}{\|\bmr^t\|_2^2} - 1\\
%   \bmx^{t+1} &= \frac{\gamma^t}{\gamma^t+1}\bmz^t\\
%   \bmr^{t+1} &= \bmy - \bH\bmx^{t+1} + \beta \bmr^t \!\frac{\gamma^t}{\gamma^t+1}\!,
%   \end{align*}
%   
% Commentted this section to make space
%%%%%%%%%%%%%%%%%%%%%%%%%%%%%%%%%%%%%%%%%%%%%%%%%%%%%%%%%%%%%%%%%
%%\subsection{Algorithm Simplifications}
%% 
%The NOPE algorithm can be simplified to:
%\begin{align*}
%\bmz^t&=\bmx^{t} + \bH^\Herm \bmr^{t}\\
%\varrho^{t} &= 1-\beta\frac{\|\bmr^t\|_2^2}{\|\bmz^t\|_2^2} \\
%\bmx^{t+1} &=\varrho^{t}\bmz^t\\
%\bmr^{t+1} &= \bmy - \bH\bmx^{t+1} + \beta \bmr^t \varrho^{t}\!,
%\end{align*}
%which only requires one division per iteration. In addition, the function $1-\beta\frac{\|\bmr^t\|_2^2}{\|\bmz^t\|_2^2}$ is well-behaved, i.e., all output values are in the range $[0,1]$; this immediately implies that $\beta\|\bmr^t\|_2^2\leq \|\bmz^t\|_2^2$ must hold. In order to prevent negative values of this function, we can compute 
%\begin{align*}
%\varrho^{t} &= \max\left\{1-\beta\frac{\|\bmr^t\|_2^2}{\|\bmz^t\|_2^2},0\right\}. 
%\end{align*}
%%%%%%%%%%%%%%%%%%%%%%%%%%%%%%%%%%%%%%%%%%%%%%%%%%%%%%%%%%%%%%%%%%%%%%%%%%%%%%%

The following result establishes the fact that NOPE achieves the same performance as that of an L-MMSE equalizer that has perfect knowledge of the signal and noise powers. %
\begin{cor} 
Let $\bH$ have uniform channel gains. In the large-antenna limit and for $t\to\infty$, NOPE as in \fref{alg:NOPE1} achieves the same SIR performance as the L-MMSE equalizer.
\end{cor}
A sketch of the proof is as follows. By following the steps in~\cite[Sec.~4.4]{mousavi2015consistent}, we see that minimizing the SURE function~\fref{eq:Psi_hat} for the NOPE algorithm results in optimal tuning.  Since optimal tuning is used, the SE equation from \fref{lem:SE} is valid for NOPE and coincides to the SIR expression for the L-MMSE equalizer given by Tse and Hanly in \cite[Thm.~3.1]{TH1999}.

\subsection{Robust Implementation of NOPE}\label{NOPE2}
NOPE, as in \fref{alg:NOPE1}, requires the channel matrix $\bH$ to have uniform gains, which is rarely satisfied in practice. We next outline how NOPE can be made robust to more general channel matrices. 
We will use ideas from the generalized approximate message passing (GAMP) algorithm~\cite{GAMP2011}. However, instead of allowing arbitrary variances in the channel matrix, we only assume that each user experiences a different variance (e.g., caused by large-scale fading).
This assumption allows us to rewrite the channel matrix as $\bH=\widetilde{\bH}\bD$, where each element of $\widetilde{\bH}$ is distributed $\setC\setN(0,1/\MR)$ and $\bD$ is a diagonal matrix containing the users' individual gains.
With this formulation, we can estimate the gain of the $\ell$-th user as
%\cs{so what is correct now???} \cj{this should be $\hat{d_\ell^2} = \sum_{j=1}^\MR \abs{H_{j,\ell}}^2$. the $1/\MR$ is embedded in $\bH$ so no need to divide the $\MR$. Original LAMA is the case when $\bD = \bI$ which agrees with $\hat{d_\ell^2} = \sum_{j=1}^\MR \abs{H_{j,\ell}}^2$.
%}
 $\hat{d_\ell^2} = \sum_{j=1}^\MR \abs{H_{j,\ell}}^2$.
Thus, $\bD$ is estimated with a diagonal matrix $\hat{\bD}$, where the $\ell$-th diagonal element is given by $\hat{d_\ell}$.
% 
%%%%%%%%%%%%%%%%%%%%%%%%%%%%%%%%%%%%%%%%%%%%%%%%%%%%%%%%%%%%%%%%%%%%%%%%%%%%%%%%%%%%%%%%%%%%%%%%%%%%%%%%
% CHARLES' version
%
%%%%%%%%%%%%%%%%%%%%%%%%%%%%%%%%%%%%%%%%%%%%%%%%%%%%%%%%%%%%%%%%%%%%%%%%%%%%%%%%%%%%%%%%%%%%%%%%%%%%%%
%
By using GAMP~\cite{GAMP2011}, we first generalize  \fref{alg:MMSE-AMP} to support nonuniform channel gains. The generalization requires us to modify the posterior mean function in \fref{eq:xhat_LMMSE} into an element-wise operation defined as 
\begin{align}\label{eq:Fmm_l}
\mathsf{F}^\text{mm}_\ell(z_\ell^t,\tau^t)= \textstyle \frac{\EX}{\EX+{\tau^t}/{\hat{d}_\ell^2}}z_\ell^t.
\end{align}
Furthermore, step \fref{eq:residual} in \fref{alg:MMSE-AMP} must be replaced by $\bmr^{t+1} = \bmy - \bH \hat{\bD}^{-2}\bmx^{t+1} + \beta \bmr^t \frac{1}{\MT} \sum_{\ell=1}^\MT \mathsf{F'}^\text{mm}_\ell(z_\ell^t,\tau^t)$.
We are now able to convert this generalized MMSE-AMP algorithm into a nonparametric algorithm by estimating the parameters~$\EX$ and $\tau^t$ in \fref{eq:Fmm_l} in the large-antenna limit as given in the following result; the proof is given in \fref{app:EX-hat}.
\begin{thm}\label{thm:EX-hat}
In the large-antenna limit, the parameters~$\EX$ and~$\tau^t$ in \fref{eq:Fmm_l} can be estimated using 
\begin{align}
\label{eq:Ex-hat}
\hat{\EX} \textstyle =\frac{\vecnorm{\hat{\bD}^{-1}\bmz^t}_2^2 - \beta \vecnorm{\bmr^t}_2^2}{\sum_{\ell=1}^\MT {\hat{d}_\ell^2}} \quad \text{and} \quad 
\hat{\tau}^t=\frac{1}{\MR}\|\bmr^t\|^2.
\end{align}
\end{thm}
\fref{thm:EX-hat} completes the necessary modifications to derive a robust version of the NOPE algorithm. 
\subsection{Numerical Results and Conclusion}
\begin{figure}
\centering
\includegraphics[width=0.78\columnwidth]{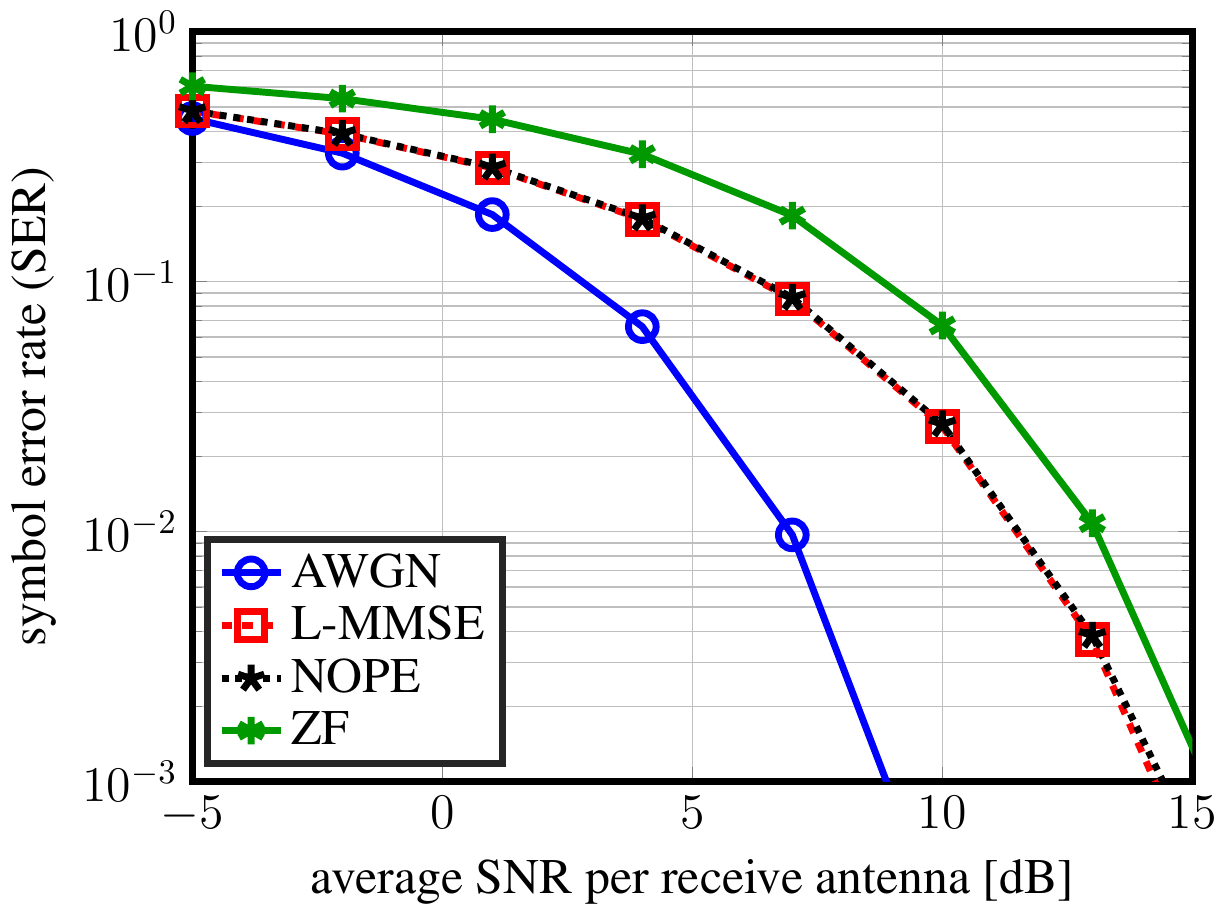}
\vspace{-0.2cm}
\caption{Symbol error-rate of NOPE algorithm in a $128\times96$ massive MU-MIMO system. NOPE closely approaches the performance of the L-MMSE estimator that requires exact knowledge of the signal  and noise power.}
\label{fig:NOPEC}
\end{figure}
\fref{fig:NOPEC} shows symbol error-rate (SER) simulation results in a $\MR=128$ and $\MT=96$ massive MU-MIMO system with QPSK modulation. 
We show the performance of ZF equalization and the AWGN lower bound as a fundamental performance limit.
Evidently, the SER performance of NOPE with 20 iterations is virtually indistinguishable from the L-MMSE estimator, which requires exact knowledge of both the signal and noise powers. Thus, NOPE is suitable for situations in which the transmit constellation may be unknown to the receiver. In addition,  NOPE often requires lower computational complexity than the L-MMSE estimator as it avoids the computation of a large matrix inversion, i.e., all involved operations for NOPE are matrix-vector products. 
%
%Hence, NOPE is not only computationally efficient but also only requires knowledge of the receive vector and the channel matrix while delivering the performance of the well-established L-MMSE equalizer. 
%
These results demonstrate that large-dimensionality of MU-MIMO systems provides the unique opportunity to design nonparametric algorithms that directly estimate the key system or model parameters from the received data, making them resilient to dynamic variations or model mismatches, while avoiding manual parameter tuning.

%\section{Conclusions}
%We have introduced NOPE, a computationally efficient data detection algorithm that does not require any knowledge regarding singal or noise parameters. NOPE achieves error-rate performance of L-MMSE in the large-antenna limit. We have established conditions for the antenna ratios under which L-MMSE performs near-optimal.
%%Broad use in regression and other estimation tasks...
%
%Aside from robustness to parameter mismatch, we use system simulations show that in realistic (finite-dimensional) systems suffering from a broad range of system impairments, NOPE (often significantly) outperforms traditional linear MMSE equalization methods that require knowledge of the signal and noise power. Furthermore, NOPE exhibits lower complexity than existing MMSE equalizers as it avoids the computation of a (typically) large matrix inversion, i.e., NOPE mainly relies on inexpensive matrix-vector products. 

\appendices
\section{Proofs}
%\subsection{Derivation of \fref{lem:MMM_SE}}\label{app:MMM_SE}
%Similarly to \cite[Sec.~\uppercase\expandafter{\romannumeral 3}-A]{JMS2016}, the mcB-AMP algorithm with the Gaussian mismatched prior $p(\bmx)=\setC \setN(0,\EXP)$ achieves the performance of a linear MMSE equalizer with mismatched signal power $\EXP$. Therefore, the SE equation for this equalizer can be obtained from the SE equation in \cite{BM2011}:
%%
%\begin{align*}
%\sigma_{t+1}^2 &= \No + \beta\min_{\gamma^2\geq0}\Exop_{X,Z}\!\left[\abs{\mathsf{F}^\text{mm}(X+\sigma_tZ,\gamma_t^2)-X}^2\right],
%\end{align*}
%where $\mathsf{F}^\text{mm}(x,\tau)=\frac{\EXP}{\EXP+\tau}x$ and the expectation is taken with respect to the true signal prior $X \sim p(x)$ and $Z\sim\setC\setN(0,1)$. 
%To simplify the SE equation, the minimization step can be computed as follows:
%\begin{align*}
%&\min_{\gamma^2\geq0}\Exop_{X,Z}\!\left[\abs{\mathsf{F}^\text{mm}(X+\sigma_tZ,\gamma_t^2)-X}^2\right]\\
%&=\min_{\gamma^2\geq0}\Exop_{X,Z}\!\left[\abs{\frac{\EXP}{\EXP+\gamma^2}(X+\sigma_tZ)-X}^2\right]\\
%&=\min_{\gamma^2\geq0}\frac{\EX \gamma^4+ {\EXP}^2 \sigma_t^2}{\left(\EXP+\gamma^2\right)^2}
%\end{align*}
%
%
%
\subsection{Proof of \fref{lem:MOR}}\label{app:MOR}
%We are looking for conditions for which the algorithm linear MMSE performs in a certain threshold of the optimal algorithm LAMA. 
%We now establish the SNR loss of MRC, ZF and L-MMSE for a given rate $R$. 
Recall that the maximum achievable rate of IO-PME at SNR $\EX/\No$ is given by $C_\text{AWGN}=\log_2(1+\EX/\No)$.
%To allow linear detectors to achieve the same rate for an SNR loss of $\DSNR$, 
%Then, assume the same system model with noise variance $\hat{\No}$. 
In the large antenna limit, the maximum achievable rates for MRC, ZF and L-MMSE at SNR $\EX/\hat{\No}$ are given by $R=\log_2\left(1+{\EX}/{\sigma^2}\right)$, where $\sigma^2$ is obtained from the following fixed-point equation:
\begin{align}\label{eq:FP_MMSE}
\sigma^2=\hat{\No}+\betaMOR \Psi(\sigma^2),
\end{align}
where $\Psi(\sigma^2)$ is $\EX$, $\sigma^2$, and $\frac{\EX}{\EX+\sigma^2}\sigma^2$ for MRC, ZF and L-MMSE, respectively~\cite{TH1999}.
Hence, the SNR loss is given by:
\begin{align*}
\nonumber
\DSNR & \textstyle = \frac{\No}{\hat{\No}}
\stackrel{\text{(a)}}{=}  \frac{\No}{\sigma^2-\betaMOR \Psi(\sigma^2)}
 \stackrel{\text{(b)}}{=}  \left(1-\betaMOR \Psi\big(\frac{\EX}{2^R-1}\big)\frac{2^R-1}{\EX}\right)^{\!-1}\!.
\end{align*}
Here, (a) follows from \fref{eq:FP_MMSE}, and (b) follows from the fact that since $R=C_\text{AWGN}$, we have $\sigma^2=\No=\frac{\EX}{2^R-1}$.
We can now extract $\betaMOR$ given $\DSNR$ as given by \fref{lem:MOR}. Moreover, for any $\beta \leq \betaMOR$ the SNR loss will be smaller than $\DSNR$.
\subsection{Proof of \fref{thm:EX-hat}}\label{app:EX-hat}

By the decoupling property of  GAMP \cite{GAMP2011},  each entry of the output $\bmz^t$  can be modeled as $z_\ell^t = \hat{d}_\ell^2 x_\ell + \hat{d}_\ell w_\ell^t$ in the large-antenna limit, with $w_\ell^t \sim \setC \setN(0,\sigma_t^2)$ where $\sigma_t^2$ is computed using the SE framework  \cite[Sec.~\uppercase\expandafter{\romannumeral 5}-C]{GAMP2011}.
% 
%Note original LAMA, it corresponds to the case where $\bD=\bI_\MT$ such that each element in $\bH$ follows $\setC\setN(0,1/\MR)$, and $\hat{d}_\ell \to 1$ so we obtain the original decoupling property.
% 
In the large-antenna limit, we estimate the signal power from the expression
\begin{align}
\nonumber
\lim\limits_{\MT\rightarrow\infty}
\textstyle\frac{1}{\MT}\vecnorm{\hat{\bD}^{-1}\bmz^t}_2^2 
& \textstyle   \stackrel{\text{(a)}}{=}
\lim\limits_{\MT\rightarrow\infty}
\!\frac{1}{\MT}\!\sum_{\ell=1}^\MT \!\Exop\!\left[\big|
z^t_\ell/\hat{d}_\ell
% \frac{z^t_\ell}{\hat{d}_\ell}
\big|^{\!2}\right]
\nonumber
\\
&\nonumber
\textstyle
\!=
\lim\limits_{\MT\rightarrow\infty}
\!\frac{1}{\MT}\!\sum_{\ell=1}^\MT {\!\Exop[|w_\ell^t|^2]\!+\! \Exop[|x_\ell|^2]  \hat{d}_\ell^2} \\
&\textstyle
= \sigma_t^2+
\hat{\EX} \lim_{\MT\rightarrow\infty}\frac{1}{\MT}\sum_{\ell=1}^\MT { \hat{d}_\ell^2}.
%\\&
%\textstyle
%\stackrel{\text{(b)}}{=}
%\lim\limits_{\MT\rightarrow\infty}\frac{1}{\MR}\vecnorm{\bmr^t}_2^2+
%\hat{\EX}\lim\limits_{\MT\rightarrow\infty}\frac{\sum_{\ell=1}^\MT {\hat{d}_\ell^2}}{\MT}.
\label{eq:Ex-derivation}
\end{align}
Here, (a) follows from Kolmogorov's strong law of large numbers~\cite{pranab1993large}  given that $\sum_{\ell=1}^\infty \!{(\EX d_\ell^2+\sigma^2)/\ell^2}\!\!<\!\! \infty$, which is satisfied since the user gains ${d_\ell^2}$ are finite for all users. 
As shown in \cite{andreaGMCS}, the parameters $\tau^t$ and $\sigma_t^2$  in \fref{eq:Ex-derivation} can be estimated from~$\|\bmr^t\|^2/\MR$.
We can finally solve \fref{eq:Ex-derivation} for $\hat{\EX}$ and obtain \fref{eq:Ex-hat}.
%
%To estimate the parameter $\tau^t$, we refer to \cite{andreaGMCS} which shows how to estimate $\tau^t$ by $\frac{1}{\MR}\vecnorm{\bmr^t}_2^2$.

% 
%Therefore, the linear estimator estimator for $z_\ell^t$ would be:
%% 
%\begin{align}\label{eq:Ffunc_GAMP}
%\mathsf{F}(z_\ell^t) &= \frac{\hat{d}_\ell^2 \EX}{\hat{d}_\ell^2 \EX + \sigma^2_t} z_\ell^t\\
%&= \frac{
%   \frac{\hat{d}_\ell^2}{\sum_{\ell=1}^\MT \hat{d}_\ell^2}\!\left(\vecnorm{\hat{\bD}^{-1}\bmz^t}_2^2 - \beta\vecnorm{\bmr^t}_2^2\right)
%}{\frac{\hat{d}_\ell^2}{\sum_{\ell=1}^\MT \hat{d}_\ell^2}\!\left(\vecnorm{\hat{\bD}^{-1}\bmz^t}_2^2 - \beta\vecnorm{\bmr^t}_2^2\right)+
%\frac{1}{\MR}\vecnorm{\bmr^t}_2^2
%}
%\\
%&= \frac{
%   \hat{d}_\ell^2\!\left(\vecnorm{\hat{\bD}^{-1}\bmz^t}_2^2 - \beta\vecnorm{\bmr^t}_2^2\right)
%}{
%\hat{d}_\ell^2\!\left(\vecnorm{\hat{\bD}^{-1}\bmz^t}_2^2 - \beta\vecnorm{\bmr^t}_2^2\right)+
%\beta\vecnorm{\bmr^t}_2^2\frac{\sum_{\ell=1}^\MT \hat{d}_\ell^2}{\MT}
%} 
%\end{align}
%
%Note that the output of the $\mathsf{F}(z_\ell^t)$ function is scaled by $\hat{d}_\ell^2$ so final $x_\ell^t$ is computed by $\mathsf{F}(z_\ell^t)/\hat{d}_\ell^2$.
% 

%\vspace{0.47cm}
\bibliographystyle{IEEEtran}
\bibliography{VIPabbrv,publishers,confs-jrnls,VIP_170121_NOPE,VIP}

\end{document}